\documentclass[11pt]{article}
\usepackage{epsf}
\usepackage{epsfig}
\usepackage{graphicx}
\usepackage{color}
\usepackage{amsmath}
\usepackage{mathrsfs}

\usepackage{amsthm}
\usepackage{latexsym}
\usepackage{amssymb}
\usepackage{amsmath}

\usepackage{stmaryrd}

\usepackage{epsf}
\usepackage{verbatim}

\newtheorem{theorem}{Theorem}[section]
\newtheorem{lemma}{Lemma}[section]

\newcommand{\nc}{\newcommand}
\nc{\Pf}{{\rm Pf}} \nc{\PfS}{{\rm PfS}}
\nc{\GP}{Grassmann-Pl{\"u}cker } \nc{\adj}{{\rm adj}}

\begin{document}
\title{A Family of Counter Examples to an Approach to Graph Isomorphism}

\vspace{0.3in}
\author{Jin-Yi Cai\thanks{University of Wisconsin-Madison, {\tt jyc@cs.wisc.edu}}
\and Pinyan Lu\thanks{Tsinghua University {\tt
lpy@mails.tsinghua.edu.cn}} \and Mingji Xia\thanks{Institute of
Software, Chinese Academy of Sciences, {\tt xmjljx@gmail.com}} }

\date{January 10,2008}
\maketitle

\bibliographystyle{plain}

\begin{abstract}
We give a family of counter examples showing that the two sequences
of polytopes $\Phi_{n,n}$ and $\Psi_{n,n}$ are different.  These
polytopes were defined recently by S. Friedland in an attempt at a
polynomial time algorithm for graph isomorphism.
\end{abstract}

\section{Introduction}
In a recent posting at arXiv (arXiv:0801.0398v1 [cs.CC] 2 Jan 2008
and arXiv:0801.0398v2 [cs.CC] 4 Jan 2008), S. Friedland defined two
sequences of polytopes $\Phi_{n,n}$ and $\Psi_{n,n}$.

Let $\Omega_n \subset {\bf R}_{+}^{n \times n}$ denote the $n \times
n$ doubly stochastic matrices. Then $\Psi_{n,n} \subset
\Omega_{n^2}$ is the convex hull of the tensor products $A \otimes
B$, where $A, B \in \Omega_n$. Meanwhile $\Phi_{n,n}$ is defined to
be the subset of $\Omega_{n^2}$ defined by the following set of
linear constraints.

\[
\sum_{j,l=1}^{n,n} c_{(i,k),(j,l)}= \sum_{j,l=1}^{n,n}
c_{(j,l),(i,k)}=1, i=1,\ldots,n,k=1,\ldots,n,\]

\[
\sum_{j=1}^{n} c_{(i,k),(j,l)}=\sum_{j=1}^{n} c_{(1,k),(j,l)},
\sum_{j=1}^{n} c_{(j,k),(i,l)}=\sum_{j=1}^{n} c_{(1,k),(j,l)},
\]
\[
\textrm{where } i=2,\ldots,n, \textrm{ and } k,l=1,\ldots,n,
\]

\[
\sum_{l=1}^{n} c_{(i,k),(j,l)}=\sum_{l=1}^{n} c_{(i,1),(j,l)},
\sum_{l=1}^{n} c_{(i,l),(j,k)}=\sum_{l=1}^{n} c_{(i,1),(j,l)},
\]
\[
\textrm{where } i=2,\ldots,n, \textrm{ and } k,l=1,\ldots,n.
\]

It was shown that $\Psi_{n,n} \subseteq \Phi_{n,n}$.  (In the
earlier version it was claimed that $\Psi_{n,n} = \Phi_{n,n}$.  If
this were the case, then graph isomorphism would be in P, as one can
reduce the problem to linear programming.  In the Jan 4th version
Friedland stated that the equality $\Psi_{n,n} = \Phi_{n,n}$ ``is
probably wrong''.) In this note we give an explicit family of
counter examples showing  $\Psi_{n,n} \not = \Phi_{n,n}$.  For every
$n \ge 4$, our examples consist of an exponential number of
matricies which are vertices of $\Phi_{n,n}$, but do not belong to
$\Psi_{n,n}$.

\section{Counter Examples}
Let $\rho  \in S_n$  be the cyclic permutation $(1 ~ 2 ~ 3 ~\ldots~
n)$. Let $\sigma \in S_n$ be any permutation.

\begin{lemma} There are exactly $n! - n \phi(n)$ many permutations
$\sigma \in S_n$, such that $\sigma \rho \sigma^{-1}$ does not
belong to the subgroup generated by $\rho$.
\end{lemma}

\begin{proof}
 A conjugate $\sigma \rho \sigma^{-1}$ of $\rho$ is also an
$n$-cycle. To be in the subgroup generated by $\rho$, iff it is a
power $\rho^i$ for some $i$ relatively prime to $n$. To be of this
form, iff $\sigma$ is of the form $\sigma(i+1) - \sigma(i)$ (in a
cyclic sense) is a constant relatively prime to $n$, which means
there are exactly $n \phi(n)$ many.
\end{proof}

Let $A$ be the matrix whose first row is $(x_1, x_2, \ldots x_n)$,
and its $i$-th row is obtained by applying $(i-1)$ times the cyclic
permutation $\rho$.  Let $B$ be the matrix whose first
 row is $(x_1, x_2, \ldots x_n)$ permuted by  $\sigma$,
and its $i$-th row is obtained by further applying $(i-1)$ times the
cyclic permutation $\rho$.

\begin{lemma} Whenever $\sigma \in S_n$ satisfies Lemma 1, there does not
exist a pair of permutation matrices $P$ and $Q$, such that $A = P B
Q$.
\end{lemma}

\begin{proof}
The first two rows of $B$ are $\sigma(x_1, x_2, \ldots x_n)$ and
$\rho\sigma(x_1, x_2, \ldots x_n)$. Assume for contradiction that
there does exist a pair of permutation matrices $P$ and $Q$, such
that $A = P B Q$. The first two rows of $B Q$ are $q \sigma(x_1,
x_2, \ldots x_n)$ and $q \rho\sigma(x_1, x_2, \ldots x_n)$, where
$q$ is the permutation corresponding to $Q$. They must be two rows
of $A$, so there exist $i$ and $j$ ($i\neq j$) such that $q
\sigma(x_1, x_2, \ldots x_n)= \rho^i (x_1, x_2, \ldots x_n)$ and $q
\rho\sigma(x_1, x_2, \ldots x_n)=\rho^j (x_1, x_2, \ldots x_n)$. We
get $\sigma^{-1} \rho \sigma=\rho ^{j-i}$, contradicting with lemma
1.
\end{proof}

Suppose $A=(a_{ij})$ is an $n \times n$ matrix. we use $\widehat{A}$
to denotes the column vector
$(a_{11},\ldots,a_{1n},a_{21},\ldots,a_{2,n},a_{3,1},\ldots,
a_{nn})^{\rm T}$ of length $n^2$.

Given $A$ and $B$, define $T$ to be the $n^2 \times n^2$ matrix
composed of $0$ and $1/n$ such that $\widehat{A}=T\widehat{B}$.

An example of this is shown as follows, for $n = 4$ and $\sigma = (3
~ 4)$:

\[
A=\left(
 \begin{array}{cccc}
x_1 & x_2 & x_3 & x_4 \\
x_2 & x_3 & x_4 & x_1 \\
x_3 & x_4 & x_1 & x_2 \\
x_4 & x_1 & x_2 & x_3
 \end{array}
 \right),
B=\left(
 \begin{array}{cccc}
x_1 & x_2 & x_4 & x_3 \\
x_2 & x_4 & x_3 & x_1 \\
x_4 & x_3 & x_1 & x_2 \\
x_3 & x_1 & x_2 & x_4
 \end{array}
 \right),
\]
\[
 T=1/4 \left(
 \begin{array}{cccccccccccccccc}
1&0&0&0&  0&0&0&1&  0&0&1&0&  0&1&0&0 \\
0&1&0&0&  1&0&0&0&  0&0&0&1&  0&0&1&0 \\
0&0&0&1&  0&0&1&0&  0&1&0&0&  1&0&0&0 \\
0&0&1&0&  0&1&0&0&  1&0&0&0&  0&0&0&1 \\
0&1&0&0&  1&0&0&0&  0&0&0&1&  0&0&1&0 \\
0&0&0&1&  0&0&1&0&  0&1&0&0&  1&0&0&0 \\
0&0&1&0&  0&1&0&0&  1&0&0&0&  0&0&0&1 \\
1&0&0&0&  0&0&0&1&  0&0&1&0&  0&1&0&0 \\
0&0&0&1&  0&0&1&0&  0&1&0&0&  1&0&0&0 \\
0&0&1&0&  0&1&0&0&  1&0&0&0&  0&0&0&1 \\
1&0&0&0&  0&0&0&1&  0&0&1&0&  0&1&0&0 \\
0&1&0&0&  1&0&0&0&  0&0&0&1&  0&0&1&0 \\
0&0&1&0&  0&1&0&0&  1&0&0&0&  0&0&0&1 \\
1&0&0&0&  0&0&0&1&  0&0&1&0&  0&1&0&0 \\
0&1&0&0&  1&0&0&0&  0&0&0&1&  0&0&1&0 \\
0&0&0&1&  0&0&1&0&  0&1&0&0&  1&0&0&0 \\
 \end{array}
 \right)
\]

\begin{theorem} For any $\sigma \in S_n$ satisfying Lemma 1.1, the matrix
$T$ is an extreme point of $\Phi_{n,n}$. However, $T \not \in
\Psi_{n,n}$.
\end{theorem}

\begin{proof}
By the definition of $A$, $B$ and $T=(t_{(i,k),(j,l)})$, for each
fixed pair $i,j$, $(t_{(i,k),(j,l)})$ (respectively, for each fixed
$k,l$, $(t_{(i,k),(j,l)})$) is a permutation matrix multiplied by
$1/n$. Obviously, $T \in \Phi_{n,n}$. For each double row index
$(i,k)$, either fix $i$, or fix $k$, and varying the other index,
and for each double column index $(j,l)$, either fix $j$, or fix
$l$, and varying the other index, we always get an $n$ by $n$
permutation matrix.

Suppose $T=\sum_s w_s T_s$, where $T_s \in \Phi_{n,n}$, $w_s>0$, and
$\sum_s w_s=1$. So within each block (fixed $i, j$, varying  $k$ and
$l$, ) the non-zero entries of $T_s$ are a subset of non-zero
entries of $T$ within that block, which form a permutation matrix.
then  by the equations for $T_s$ within the block,  it must be
either totally zero or a positive multiple of the same permutation
matrix made up of non-zero entries of $T$ within that block.  For
each block, the permutation matrix is the same for every $T_s$. The
multipliers form a doubly stochastic matrix $M_s \in \Omega_n$, by
the global sum $\sum_{j,l=1}^{n,n} = 1$. Therefore  $T_s$ is as
follows: its $(i,j)$ block is obtained by multiplying each entry of
a doubly stochastic matrix $M_s \in \Omega_n$ with the permutation
matrix of $T$ for each block.

Now if we consider the sum $\sum_{j=1}^{n} c_{(i,k),(j,l)} =
\sum_{j=1}^{n} c_{(1,k),(j,l)}$, by the property of $T$ each row of
$M_s$ is a constant.  (Similarly each column of $M_s$ is a
constant.) Thus $M_s$ is just the all $1/n$ matrix $1/n J$.

This implies that there is exactly one term in the sum $T = \sum_s
w_s T_s$, and $T$ is an extreme point.

Assume for a contradiction that $T \in \Psi_{n,n}$ and $T=\sum_s w_s
P_s \otimes Q_s$, where $P_s, Q_s$ are permutation matrices,
$w_s>0$, and $\sum_s w_s=1$. We get $T\geq w_1 P_1 \otimes Q_1$
(Here the relation of $\geq$ is entry-wise). For any $x_1, x_2,
\ldots, x_n \geq 0$, $T \widehat{B}\geq w_1 P_1 \otimes Q_1
\widehat{B}$, that is, $A\geq w_1 P_1 B Q_1$. By lemma 1.2, $P_1 B
Q_1$ is different from $A$, so there must be an entry $(i,j)$ such
that they are different at that entry. Notice that each entry of $A$
or $P_1 B Q_1$ is a single variable from $\{x_1,\ldots, x_n\}$.
W.l.o.g, we can assume the $(i,j)$-th entry of $A$ and $P_1 B Q_1$
are $x_1$ and $x_2$. We can set $x_1=0$ and $x_2=1$ such that
$A_{ij}< (w_1 P_1 B Q_1)_{ij}$, which is a contradiction.  So $T
\not \in \Psi_{n,n}$.
\end{proof}

Before we posted this note, we note that Babai \\
(http://people.cs.uchicago.edu/~laci/polytope.pdf) and Onn
(arXiv:0801.1410) have both pointed out that the linear optimization
problem over the polytope $\Psi_{n,n}$ can solve NP-complete
problems, and therefore it is unlikely that $\Psi_{n,n}$ can be
defined by a polynomial number of
(in)equalities as $\Phi_{n,n}$ can.  In\\
(http://people.cs.uchicago.edu/~laci/polytope-correspondence.pdf),
Babai also mention that Joel Rosenberg already gave a counter
example showing the two polytopes are different, for $n=4$.

\end{document}